\documentclass[12pt,preprint]{aastex}
\begin{document}
\title{The Variation of Galaxy Morphological Type with the Shear of 
Environment} 
\author{Jounghun Lee and Bomee Lee}
\affil{Department of Physics and Astronomy, FPRD, Seoul National University,
Seoul 151-747, Korea}
\email{jounghun@astro.snu.ac.kr, bmlee@astro.snu.ac.kr}
\begin{abstract}
Recent N-body simulations have shown that the assembly history of galactic 
halos depend on the density of large-scale environment. It implies that the 
galaxy properties like age and size of bulge may also vary with the 
surrounding large-scale structures, which are characterized by the tidal shear 
as well as the density. By using a sample of $15,882$ well-resolved nearby 
galaxies from the Tully Catalog and the real space tidal field reconstructed 
from the 2Mass Redshift Survey (2MRS), we investigate the dependence of galaxy 
morphological type on the shear of large-scale environment where the galaxies 
are embedded. We first calculate the large scale dimensionless overdensities 
($\delta$) and the large-scale ellipticities ($e$) of the regions where the 
Tully galaxies are located and classify the Tully galaxies according to their 
morphological types and create subsamples selected at similar value of 
$\delta$ but span different ranges in $e$. 
We calculate the mean ellipticity, $\langle e\rangle$,  averaged over each 
subsample and find a signal of variation of $\langle e\rangle$ with galaxy 
morphological type: For the case of $0.5\le\delta\le1.0$, the ellipticals are 
found to be preferentially located in the regions with low ellipticity. 
For the case of $-0.3\le\delta\le 0.1$, the latest-type spirals are found 
to be preferentially located in the regions with high ellipticity. 
The null hypothesis that the mean ellipticities of the regions where the 
ellipticals and the latest type spirals are located are same as the global 
mean ellipticity averaged over all types is rejected at $3\sigma$ level 
when $-0.3\le\delta\le 0.1$. Yet, no signal of galaxy-shear correlation is 
found in the highly overdense/underdense regions. The observed trend suggests 
that the formation epochs of galactic halos might be a function not only of 
halo mass and large-scale density but also of large-scale shear. 
Since the statistical significance of the overall trend is low, it will 
require a sample of at least $100,000$ galaxies to verify the existence of  
this correlation.
\end{abstract}
\section{INTRODUCTION}

It has been known for long that the physical properties of galaxies such as 
morphological type, color, luminosity, spin parameter, star formation rate, 
concentration parameter and so on are functions of their environments 
\citep{dre80,pos-gel84,whi-etal93,lew-etal02,gom-etal03,got-etal03,roj-etal05,
kue-ryd05,bla-etal05,ber-etal06,cho-etal07,par-etal07}. 
Most previous works have largely focused on the galaxies located in the highly 
dense regions and usually quantified the environmental dependence of galaxy 
properties in terms of cross-correlations with local density on small scale, 
which is  believed to be established by environment-dependent processes like 
galaxy-galaxy interaction.

The currently popular $\Lambda$CDM model predicts that the galaxy properties 
are correlated with not only the small scale but also the large-scale 
environment. Here the small scale and the large-scale environments represent 
the surrounding regions smaller and larger than $5h^{-1}$Mpc, respectively.
Recent high-resolution N-body simulations of a $\Lambda$CDM cosmology have 
demonstrated that the assembly history of galactic halos of a given mass 
change with the density field smoothed on sufficiently large scale 
\citep{gao-etal05}. Since the galaxy content depend strongly on assembly 
history of its host halo \citep{spr-etal05,cro-etal07}, this numerical 
finding suggests that the intrinsic properties of galaxies may be correlated 
with the density of large-scale environments. 

In fact, several groups have already found observational evidences for the 
existence of cross-correlations between galaxy properties and large-scale 
tidal shear. \citet{nav-etal04} found that the spin axes of spiral galaxies 
near the Local Supercluster lie preferentially on the Supergalactic Plane. 
\citet{tru-etal06} also showed by analyzing recent observational data from 
large galaxy surveys that the spin axes of void galaxies tend to be inclined 
to the void surfaces, keeping the initial memory of tidally induce alignments. 
\citet{pan-bha06} have shown that the galaxy luminosity and colour are 
dependent on the filamentarity of large-scale environment. More recently, 
\citet{pan-bha08} have also presented that the star formation rate of galaxies 
are cross-correlated with the filamentarity of the surrounding large-scale 
structures. \citet{her-etal07} found a one-to-one correspondence between 
the value of the galaxy spin parameter and the morphological type, while 
\citet{her-cer06} and \citet{cer-her08} have noted that the spin parameter 
of the observed galaxies is strongly correlated with their color and 
chemical abundance.

The existence of cross-correlations between galaxy properties and density of 
large-scale environment implies that the galaxy content still have the memory 
of the initial conditions of the Lagrangian regions where the host halos 
formed. In the scenario of the biased galaxy formation, 
the galaxy sites correspond to the high peaks of the linear density field 
\citep{kai84,bar-etal86}. But, since the initial density peaks are not 
spherical, they should be characterized not only by the peak heights 
but also by their ellipticity and prolateness. Hence, if the galaxy 
properties are still linked to the large-scale density, then it is likely 
that they are also linked to the large-scale ellipticity and prolaticity.

Our goal here is to test this statistical link. Instead of the ellipticity 
of the large-scale density field, however, we consider the ellipticity of 
the large-scale potential field (i.e., the tidal shear). Since  the potential 
field is smoother than the density field, we believe that the large-scale 
potential field reflects more directly the linear conditions. 
Furthermore, the tidal shear field, defined as the second derivative of the 
potential field, has a large-scale coherence, resulting in highly anisotropic 
spatial distribution of galaxies at present epoch \citep{bon-etal96}. 
Therefore, the tidal shear field is more readily measurable from the 
large-scale spatial distribution of present galaxies.

In previous approaches, however, the correlation between galaxy properties 
and the shear of large-scale structure was measured in redshift space. 
Since the redshift distortion effect could contaminate the measurements of 
the filamentarity of the large scale structure, it will be desirable to 
measure it in real space. Moreover, true as it is that in the linear regime 
the density and the tidal shear at a given region are mutually independent, 
it is likely that the two quantities have developed cross-correlations in the 
subsequent non-linear regime. Therefore, to find an independent relationship 
between galaxy properties and large-scale tidal shear, one has to first 
account for the density-shear correlations and remove its effect.
 
We attempt here to measure observationally an independent relationship 
between the morphological types of nearby large galaxies and the real-space 
tidal shear reconstructed from all sky survey. The organization of this paper 
is as follows:  In \S 2, the observational data are described.  
In \S 3, the shear of environment is defined and its mean value averaged over 
each galaxy subsample at similar density is measured as a function of galaxy 
morphological type. In \S 4, the results are discussed and a final conclusion 
is drawn.

\section{OBSERVATIONAL DATA: AN OVERVIEW}

A total of $35,000$ galaxies with spectroscopic information are listed in the 
Tully Catalog, which was constructed from the ESO/Uppsala full-sky survey 
\citep{nil74,lau82}.  Only $15,922$ galaxies in the Tully catalog have  
detailed information on their morphological types, which were determined 
by B. Tully according to the conventional Hubble classification scheme. 
For a detailed description of the Hubble classification scheme, see Table 2  
in the Third References Catalog of Bright Galaxies \citep[RC3][]{dev-etal91}. 
As for the rest $19,078$ galaxies in the Tully catalog, the stages of their 
morphological types are described as either "irregulars" or "uncertain" or 
"doubtful" or "spindle", "outer ring" or "pseudo-outer R", or "irregulars".  
These are thus excluded from our analysis.

According to the morphological type described in RC3, we divide the selected 
Tully galaxies into six samples: E, L, SI, SII, SIII, and SIV. 
Table \ref{tab:type} lists the class, the morphological type (as in RC3), 
and the number of the galaxies ($N_{g}$) belonging to the six samples: 
The sample "E" contains only ellipticals with the Hubble type of cE,E0, E0-1, 
E+; The sample $L$ contains only lenticulars with the Hubble type of 
$S0^{-1}$,$S0^{O}$,$S0^{+}$; The sample $SI$ consists of the spirals of types 
$S0/a$ and $Sa$ which have tightly wound arms and a largest bulge. 
While the spirals of types $Scd$, $Sd$, $Sdm$ and $Sm$ belong to the sample 
$SIV$, which have no well developed bulge and loosely wound arms. 
The sample $SII$ contains the spirals of type $Sa$ as well as the spirals 
of type $Sab$ that are intermediate between $Sa$ and $Sb$. 
Likewise, the sample $SIII$ contains the spirals type of $Sb$ and 
the spirals of type $Sbc$. As can be noted, the size of bulge and the 
tightness of the spiral arms decrease from $SI$ to $SIV$. 

The real space density field was originally reconstructed on $64^{3}$ grids 
in a box of linear size $400h^{-1}$Mpc from the 2Mass Redshift Survey 
\citep[][2MRS]{erd-etal06}. Basically, it is the linear density field 
smoothed with a wiener filter assuming a linear bias. \citet{lee-erd07} used 
the 2MRS density field to reconstruct the tidal shear field on the same 
$64^{3}$ grids. As described in detail in \citet{erd-etal06}, the real space 
positions were recovered by applying the Wiener reconstruction algorithm to 
the 2MRS data. This algorithm basically deconvolve the linear redshift 
distortions in the radial direction using a distortion matrix. 
For a detailed description of the method used to recover the real-space 
position, see \citet{erd-etal06}.

\begin{figure}
\begin{center}
\plotone{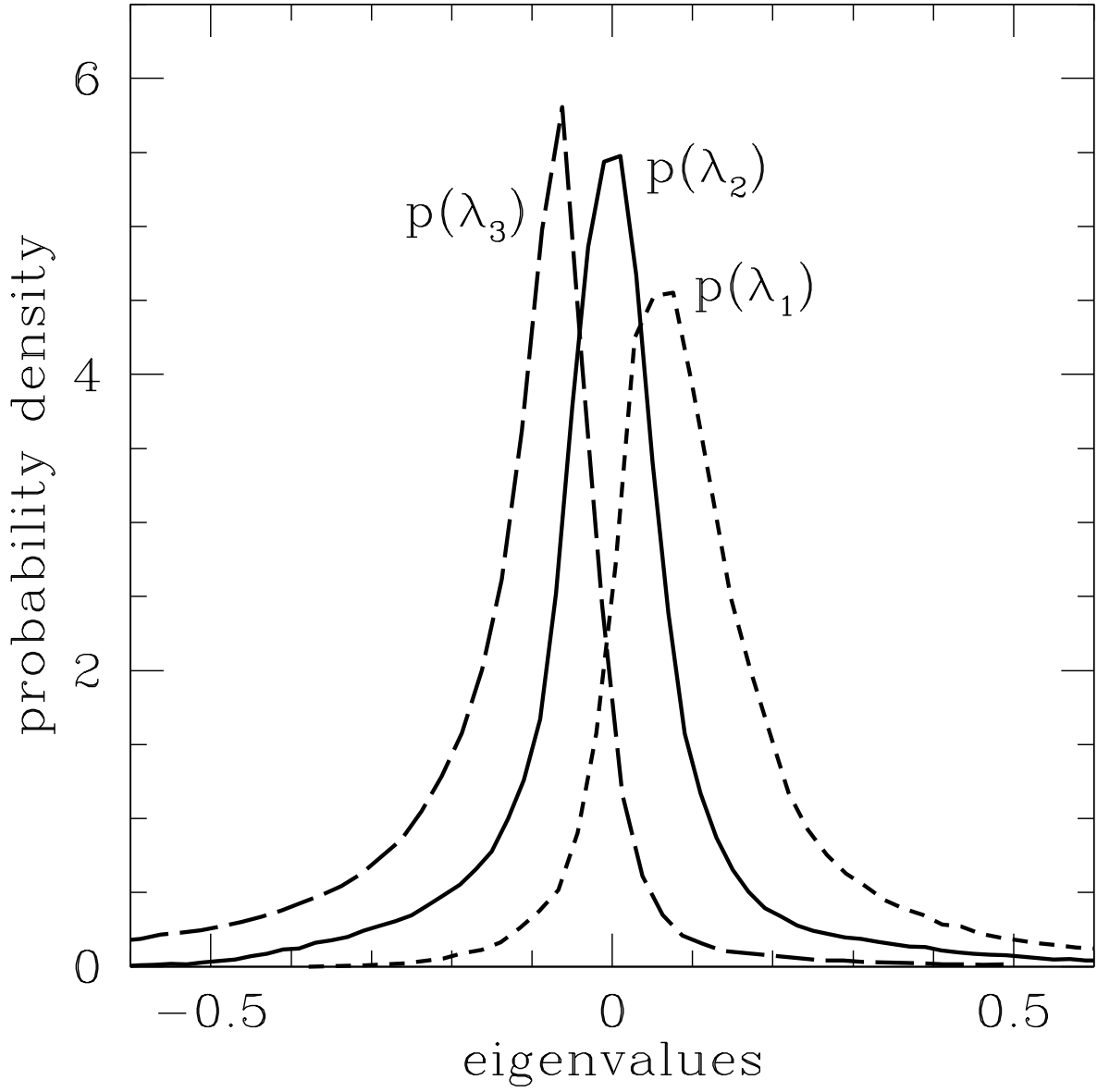}
\caption{The probability density distributions of the three eigenvalues 
of the 2MRS tidal shear field.}
\label{fig:pro}
\end{center}
\end{figure}

The reconstructed tidal shear field consists of a set of three eigenvalues, 
$\{\lambda_{1},\lambda_{2},\lambda_{3}\}$, of the tidal shear tensor assigned 
to each grid. Figure \ref{fig:pro} plots the probability density distributions 
of the three eigenvalues of the 2MRS tidal field, assuming 
$\lambda_{1}\ge\lambda_{2}\ge\lambda_{3}$.  As can be seen, the distribution 
of the second largest eigenvalue, $p(\lambda_{2})$, is almost symmetric around 
the peak at zero with narrow width, while the other two distributions 
$p(\lambda_{1})$ and $p(\lambda_{3})$ are asymmetric around zero with 
broader shapes. 

\begin{figure}
\begin{center}
\plotone{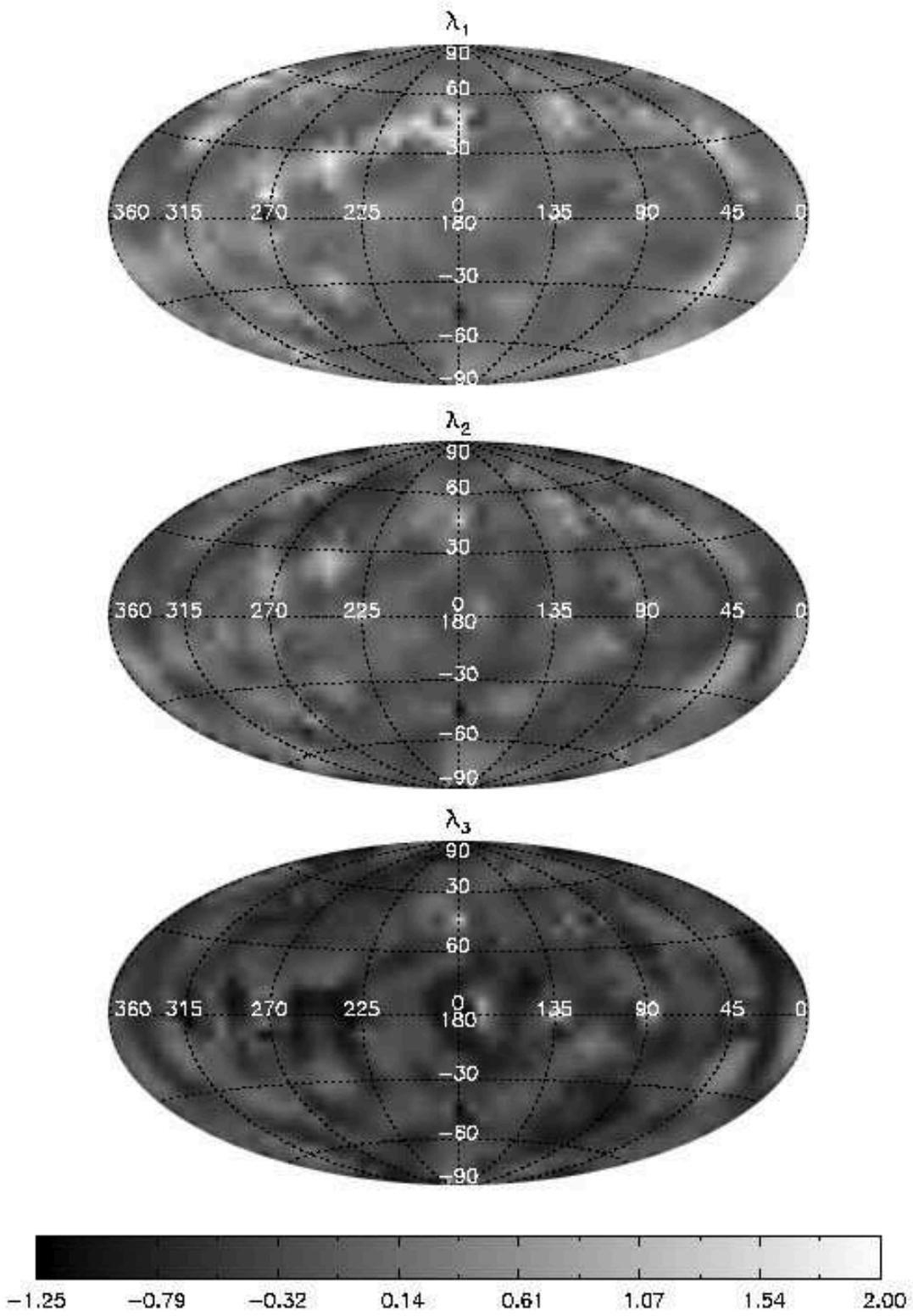}
\caption{The Aitoff Projections of the tidal shear eigenvalue field in 
Supergalactic coordinates at $8000$ km/s: $\lambda_{1}$ (top), $\lambda_{2}$ 
(middle), $\lambda_{3}$ (bottom)}
\label{fig:ait}
 \end{center}
\end{figure}

The differences among the eigenvalues are related to the shear of environment 
at a given grid. If $\lambda_{1}=\lambda_{2}=\lambda_{3}$, it means that the 
shear of environment is zero. The larger the mutual differences among the 
eigenvalues are, the higher the shear of environment is. Figure \ref{fig:ait} 
plots the 2MRS tidal shear eigenvalue field evaluated on a thin shell at 
$80h^{-1}$Mpc shown in Supergalactic Aitoff projection. The top, middle 
and bottom panel shows the $\lambda_{1}$, $\lambda_{2}$ and $\lambda_{3}$ 
field, respectively. In each panel, the darker regions correspond to 
the low eigenvalues while the bright regions to the high eigenvalues. 

By applying the cloud-in-cell (CIC) interpolation method \citep{hoc-eas88} to 
the 2MRS tidal shear field, \citet{lee-erd07} also calculated the tidal 
tensors and their eigenvalues at the positions of the selected $15,922$ 
Tully galaxies. Using the reconstructed 2MRS tidal shear field, we measure 
the values of the shear at the positions of the Tully galaxies.  In \S 3, 
we will see how this measured shears are related to the morphological types 
of the Tully galaxies.

\section{Galaxy-Shear Correlations}

\subsection{\it Variation with the LSS Type}

Let us consider a position, ${\bf x}$, where a selected Tully galaxy is 
located. It depends on the signs of $\lambda_{1},\ \lambda_{2},\ \lambda_{3}$ 
at ${\bf x}$, whether the selected galaxy belongs to a void or a sheet or 
a filament or a halo. If all three eigenvalues are positive at ${\bf x}$, 
then the given galaxy is located in a halo-like region; If $\lambda_{2}>0$, 
$\lambda_{3}<0$, then it is in a filament-like region; If $\lambda_{1}>0$, 
$\lambda_{2}<0$, then it is in a sheet-like region; If $\lambda_{1}<0$, 
then it is in a void-like region. 

We investigate the dependence of the relative abundance of the sample galaxies 
located in halo-like regions on the morphological type, by measuring the 
conditional number density (CND):
\begin{equation}
\frac{\delta N_{\rm type}}{\delta N_{\rm all}} 
\equiv \frac{\Delta N_{\rm type}}{N_{\rm type}}\times
\frac{N_{\rm all}}{\Delta N_{\rm all}}.
\end{equation}
Here, $N_{\rm type}$ is the number of the galaxies belonging to a given 
sample, $\Delta N_{\rm type}$ is the number of those galaxies in the sample 
which are located in halo-like regions, $N_{\rm all}$ is the number of 
all selected Tully galaxies and $\Delta N_{\rm all}$ is the number of those 
Tully galaxies which are located in halo-like regions. If CND $>1$, then 
the galaxies in a given sample have stronger tendency to be located in 
halo-like regions than the parent sample. 

\begin{figure}
  \begin{center}
   \plotone{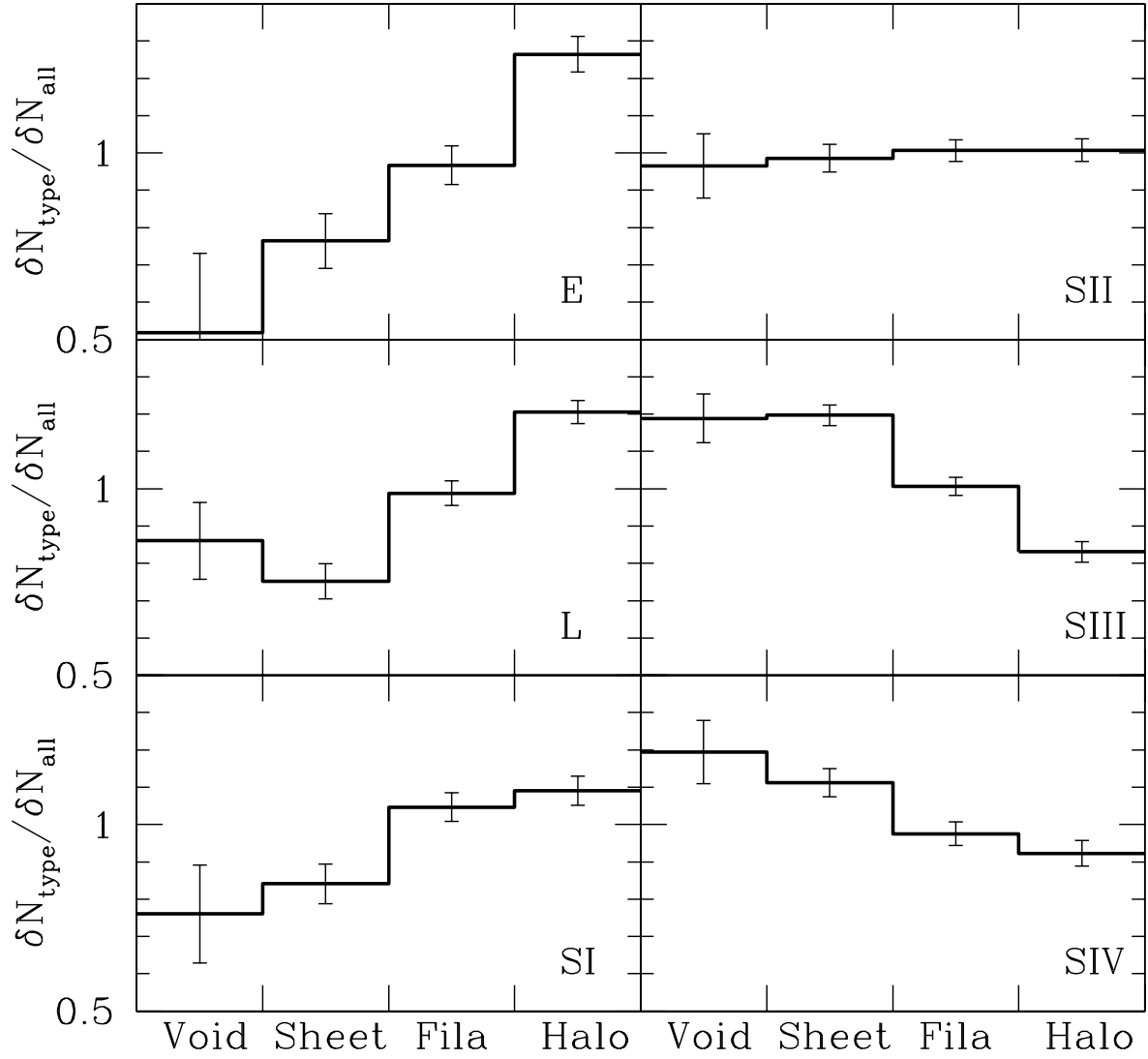}
\caption{The conditional number densities of the galaxies located in the 
void-like, sheet-like, filament-like and halo-like regions for the six 
subsamples with Poisson errors.}
\label{fig:con}
 \end{center}
\end{figure}

The CND of the galaxies located in void-like, sheet-like, filament-like 
regions are also calculated in a similar manner. Figure \ref{fig:con} plots 
the CNDs as histograms for the six samples with Poisson errors.  
As can be seen, the six histograms show all distinct behaviors: 
The CND of the sample E is highest in halo-like regions and lowest in 
void-like regions; The CND of the sample L is lowest in sheet-like regions; 
The CND of the sample SI is highest in  filament-like regions; 
The CND of the sample SII is almost uniform; The CND of the sample SIII is 
highest in void-like and sheet-like regions; The CND of the sample SIV is 
highest in void-like regions. It is interesting to see that in void-like 
regions the CND of the lenticulars have relatively high value compared with 
that of the ellipticals. 

Although the result shown in Fig. \ref{fig:con} shows clearly that the galaxy 
morphology is correlated with the shear of the surrounding large-scale 
structure, it should not be directly translated into an observational evidence 
for the existence of an {\it independent} relationship between galaxy 
morphology and large-scale shear. 
Since the signs of the three eigenvalues of the tidal shears are correlated 
with the local density, the observed signals shown in Fig. \ref{fig:con} 
could be resulted from the well known correlations between the local density 
and the galaxy morphology. For instance, the observed tendency that the 
ellipticals are more abundant in the halo-like regions may be due to the 
fact that the local density is usually higher in the halo-like regions 
since the three eigenvalues are all positive and the ellipticals are 
preferentially located in the high-density regions. Therefore, to detect the 
true relationship between galaxy morphology and the shear of large-scale 
structure, one has to use controlled subsamples where the correlations between 
galaxy morphology and density are removed. In the following subsection, we 
pursue this task.

\subsection{\it Variation with the Ellipticity}
 
The shear of environment at a given galaxy position ${\bf x}$ is caused by the 
anisotropy in the distribution of the surrounding large scale structure which 
in turn induces asphericity in the gravitational potential $\Phi ({\bf x})$. 
The asphericity of $\Phi ({\bf x})$ can be quantified in terms of ellipticity, 
$e$, defined as \citep{bar-etal86}
\begin{equation}
e\equiv\frac{\lambda_{1}-\lambda_{3}}{2\vert\delta\vert}.
\label{eqn:ep}
\end{equation}
Here the sum of the three eigenvalues of the tidal shear equals the 
dimensionless overdensity, $\delta \equiv (\rho - \bar{\rho})/\bar{\rho}$ 
where $\bar{\rho}$ is the background density: 
$\delta=\sum_{i=1}^{3}\lambda_{i}$. 
If the three eigenvalues, $\lambda_{1},\ \lambda_{2},\ \lambda_{3}$, at 
${\bf x}$ have the same value, the iso-potential surface at ${\bf x}$ has 
a spherical shape with $e=0$. The larger the differences between 
$\lambda_{1}$ and $\lambda_{3}$ are, the higher the value of $e$ is. 
If the spatial distribution of galaxies located in a region is highly 
anisotropic, then the value of $e$ of the region will be high. In other 
words, the region has a high ellipticity. The higher the degree of the 
anisotropy in the spatial distribution of the galaxies in a region increase, 
the higher value of the ellipticity the region has.

\begin{figure}
  \begin{center}
   \plotone{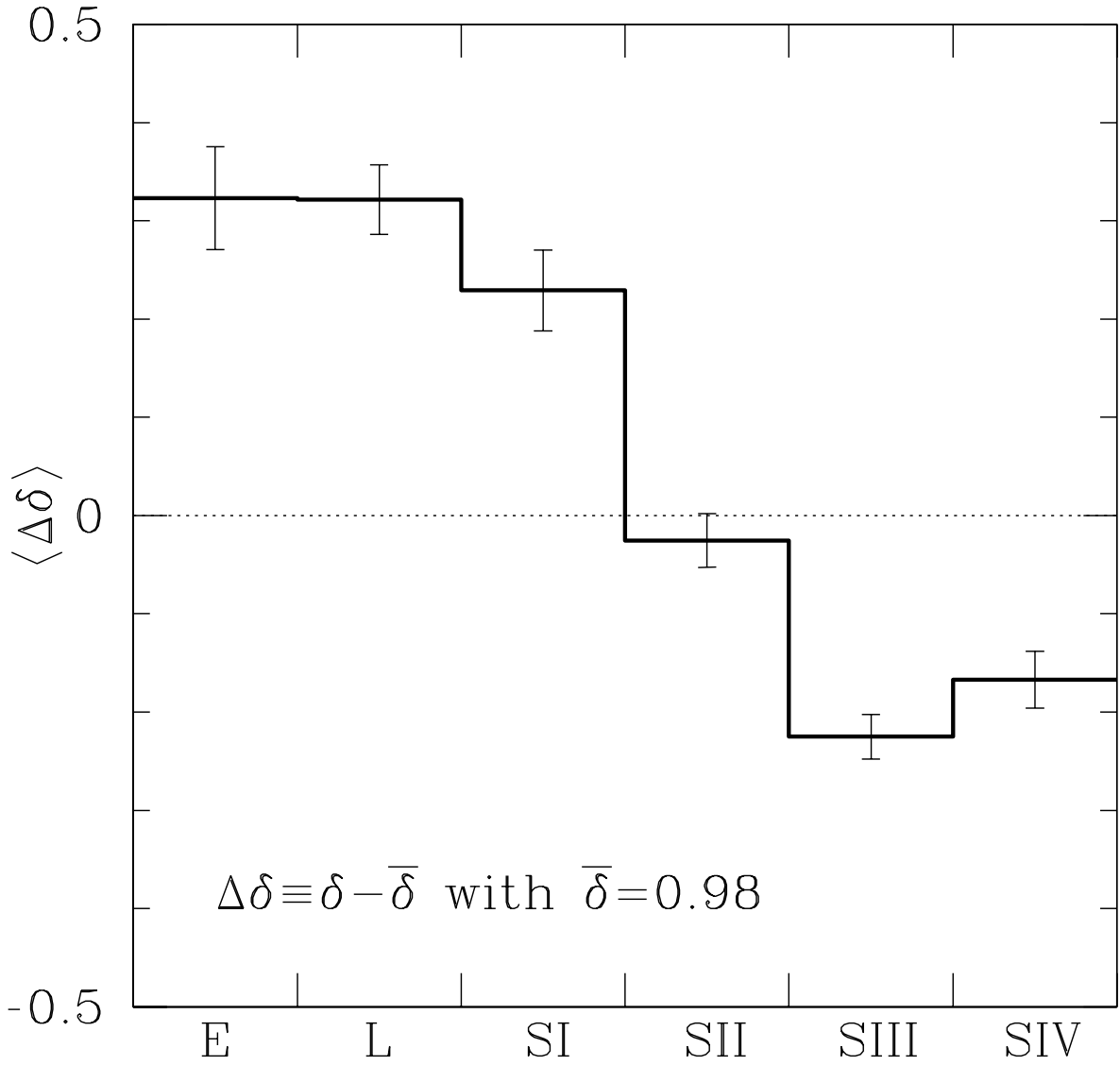}
\caption{Mean of the density difference averaged over the regions where the 
Tully galaxies of each subsample are located. The Tully galaxies are 
classified into six subsamples according to the galaxy morphological type. 
The errors represent one standard deviation in the measurement of the 
mean values.}
\label{fig:den}
 \end{center}
\end{figure}

At the position of each Tully galaxy, we determine the values of $\delta$ 
and $e$. We first calculate the means of $\Delta\delta\equiv\delta
-\bar{\delta}_{G}$ averaged separately over each sample listed in Table 
\ref{tab:type}, where $\bar{\delta}_{G}$ is the global mean value of the 
dimensionless overdensity averaged over all selected Tully galaxies. 
If $\Delta\delta >0$, there is a tendency of the sample galaxies to be 
located in high-density regions. If $\Delta\delta <0$, the sample galaxies 
tend to be located in low-density regions. 
If $\Delta\delta =0$ for all six samples, there is no correlation 
between galaxy morphology and large-scale density. 
Figure \ref{fig:den} plots $\langle\Delta\delta\rangle$ of the six samples.
As can be seen, there is indeed a strong dependence of galaxy-morphology 
on the large-scale density, which is consistent with numerical results 
\citep{gao-etal05}.

Now, we calculate the means of $\Delta e\equiv e-\bar{e}_{G}$ averaged 
separately over each sample, under the constraint that the value of $\delta$ 
is fixed in some narrow range to remove the effect of morphology-density 
correlations. Note that $\bar{e}_{G}$ is averaged in the same constrained 
range of $\delta$. If $\Delta e >0$, there is a {\it independent} tendency 
of the sample galaxies to be located in the high-shear regions. 
If $\Delta e <0$, the sample galaxies tend to be located in the low-shear 
regions. If $\Delta e=0$ for all six samples, there is no correlation between 
galaxy morphology and large-scale shear. 

\begin{figure}
  \begin{center}
   \plotone{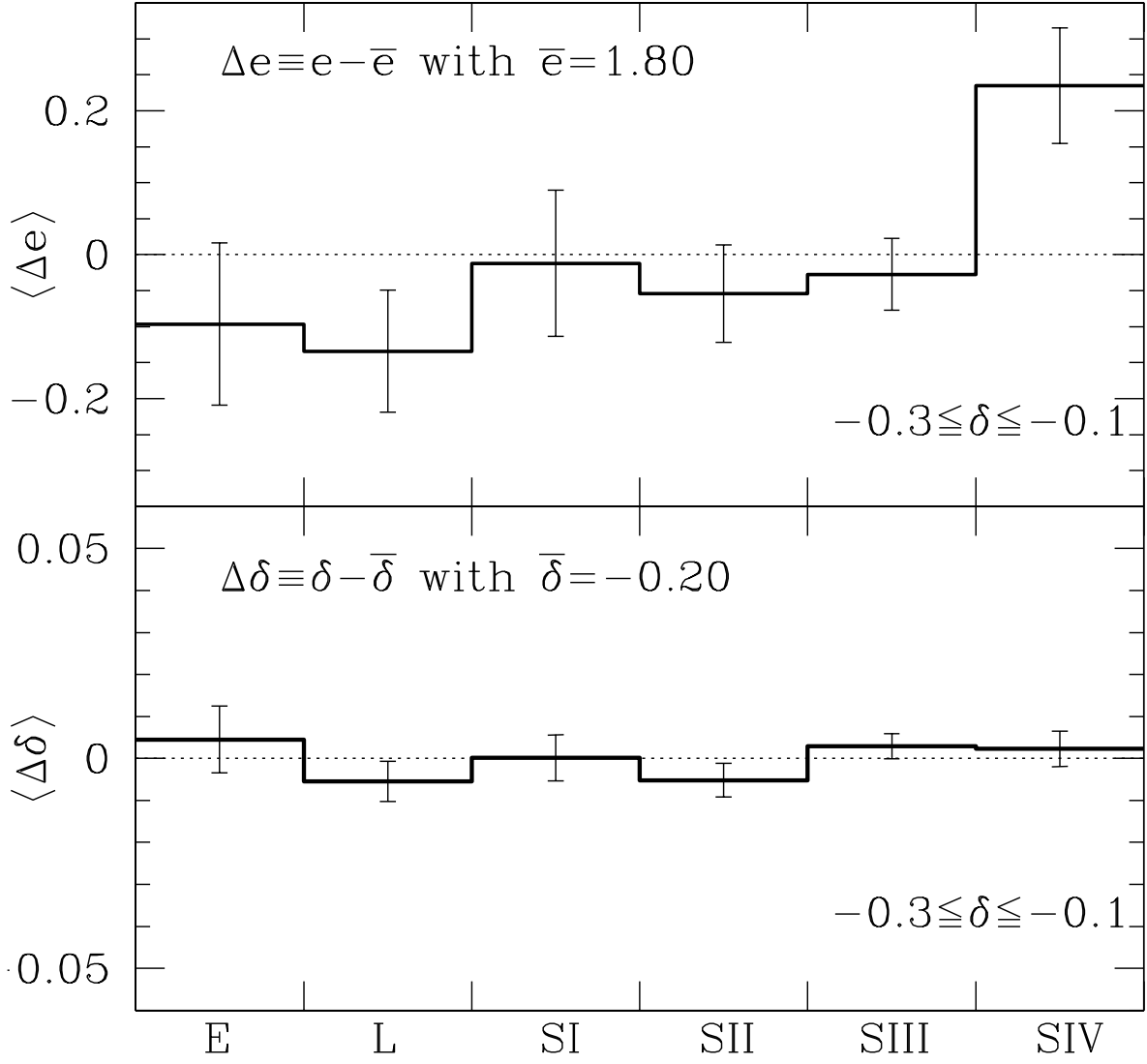}
\caption{Mean of the ellipticity difference (top) and the density 
difference (bottom) averaged over each subsample with $\delta$ in the 
range of $[-0.3,-0.1]$.}
\label{fig:ell1}
 \end{center}
\end{figure}

In Fig. \ref{fig:ell1} we plot $\langle\Delta e\rangle$ of the six samples in 
the top panel for the case where the value of $\delta$ is fixed in the range 
of $[-0.3,-0.1]$.  The errors, $\sigma$, are calculated as one standard 
deviation in the measurement of the mean value: 
$\sigma\equiv\sqrt{[\langle(\Delta e)^{2}\rangle-\langle\Delta e\rangle^{2}]
/(N_{g}-1)}$ where $N_{g}$ is the number of the sample galaxies selected 
at the given density range. To demonstrate that the correlation between 
morphology and density is effectively removed by constraining the value of 
$\delta$ to this narrow range of $[-0.3,-0.1]$, we also plot 
$\langle\Delta\delta\rangle$ of the six samples in the bottom panel. 
As can be seen, the value of $\langle\Delta\delta\rangle$, is almost uniform 
over the six subsamples, indicating that the morphology-density correlations 
are controlled to a  negligible level with this constraint of 
$-0.3\le\delta\le-0.1$. We detect a $2.9\sigma$ signal that the mean 
ellipticity of the sample $SIV$ deviates from the global mean ellipticity. 
This result suggests that the latest-type spirals tend to be located in the 
regions of high-ellipticity. 
 
\begin{figure}
  \begin{center}
   \plotone{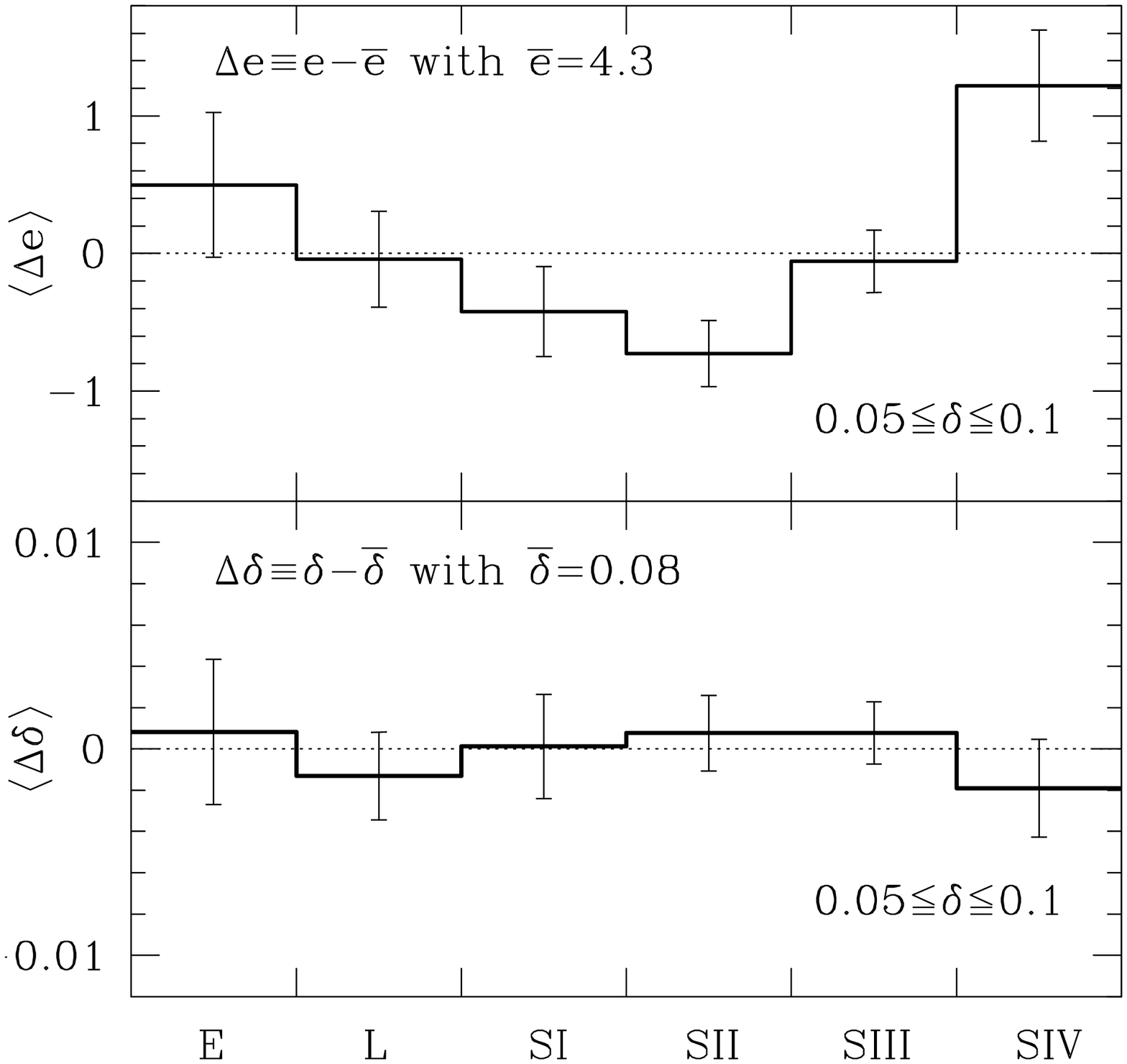}
\caption{Same as Fig. \ref{fig:ell1} but with $\delta$ in the 
range of $[0.05,0.1]$.}
\label{fig:ell2}
 \end{center}
\end{figure}

\begin{figure}
  \begin{center}
   \plotone{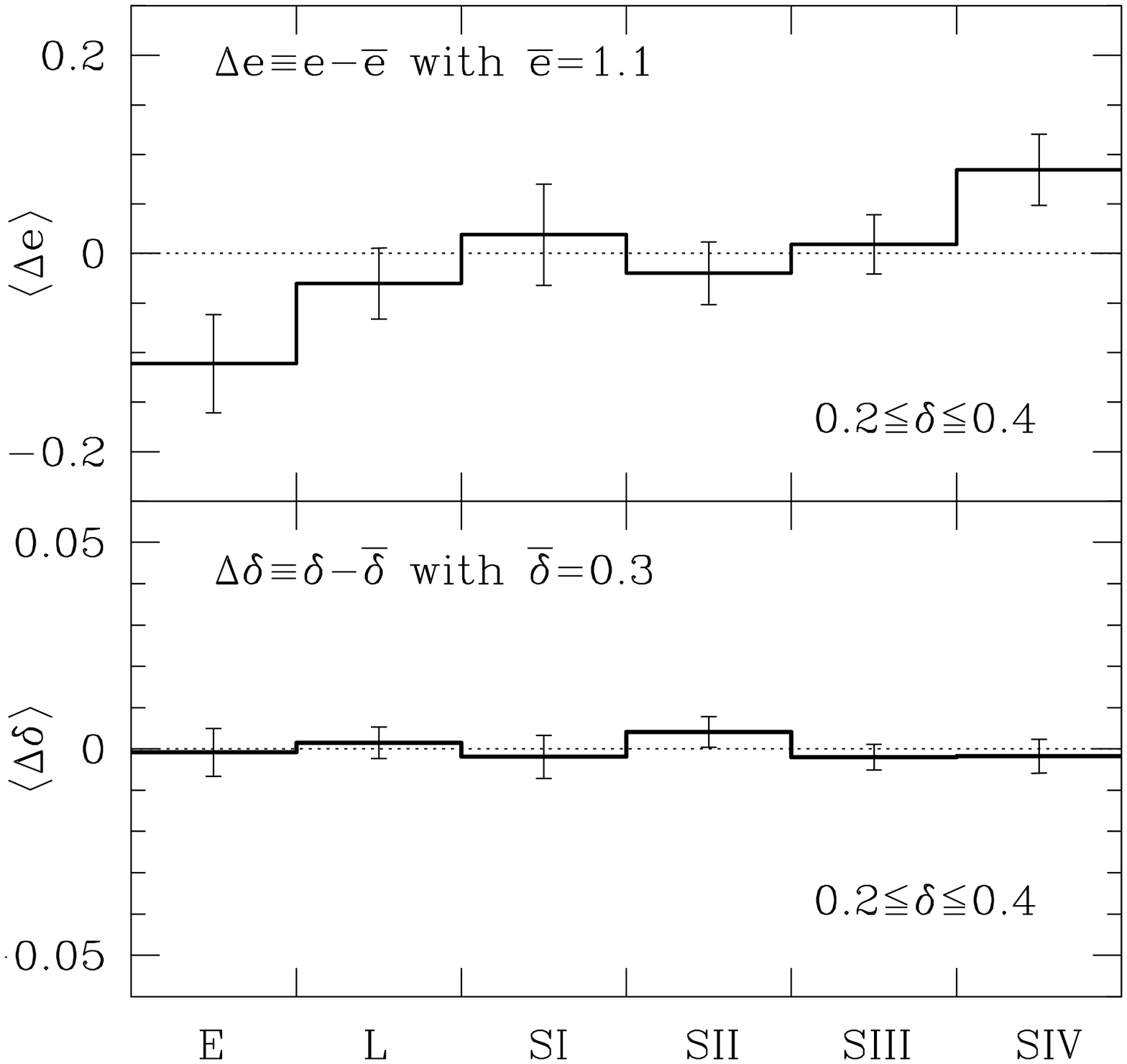}
\caption{Same as Fig. \ref{fig:ell1} but with $\delta$ in the 
range of $[0.2,0.4]$.}
\label{fig:ell3}
 \end{center}
\end{figure}

\begin{figure}
  \begin{center}
   \plotone{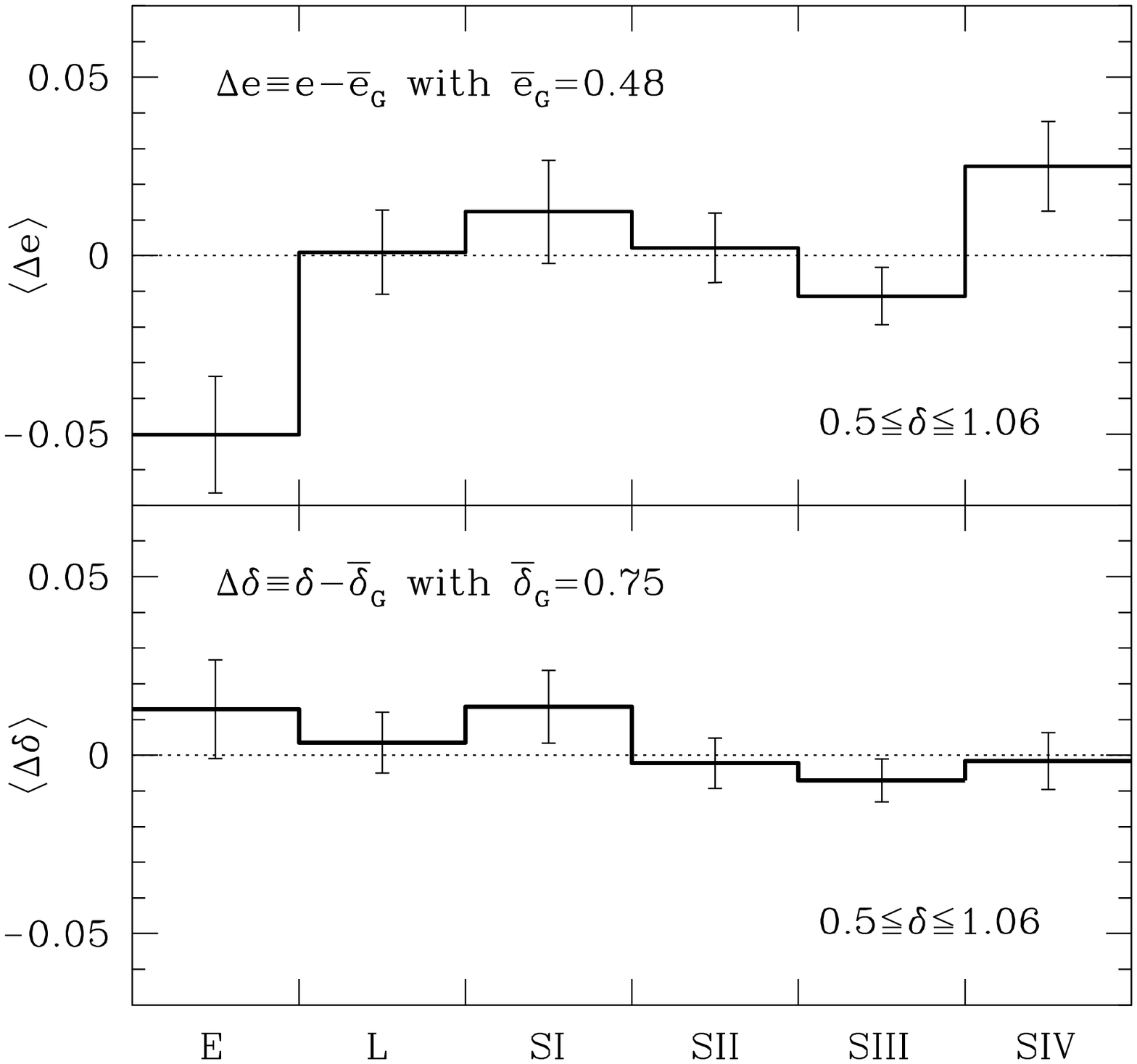}
\caption{Same as Fig. \ref{fig:ell1} but with $\delta$ in the 
range of $[0.5,1.06]$.}
\label{fig:ell4}
 \end{center}
\end{figure}

Fig. \ref{fig:ell2} plots the same as Fig. \ref{fig:ell1} but with a different 
constraint of $0.05\le\delta\le 0.1$. A $3\sigma$ signal of morphology-shear 
correlation is also detected for the sample $SIV$ galaxies. It is consistent 
with the previous result that the latest-type spirals tend to be 
preferentially located in the high-shear environment. 
Fig. \ref{fig:ell3} plots the same as Fig. \ref{fig:ell1} but with $\delta$ 
in the range of $[0.2,0.4]$.  We detect a $2.2\sigma$ signal that the 
mean ellipticity of the regions where the galaxies of the sample $E$ are 
located is lower than the global mean ellipticity, and a $2.4\sigma$ signal 
that the mean ellipticity of the regions where the galaxies of the sample 
$SIV$ is located is higher than the global mean ellipticity.   
Fig. \ref{fig:ell4} plots the same as Fig. \ref{fig:ell1} but with $\delta$ 
in the range of $[0.5,1.06]$. There are detected a $3.1\sigma$ signal that the 
mean ellipticity of the regions where the sample $E$ galaxies are located 
is lower than the global mean ellipticity, and a $2\sigma$ signal of 
correlation between latest type spirals (SIV) and high-shear region. 
The results shown in Figs. \ref{fig:ell3} and \ref{fig:ell4} indicate 
consistently that the ellipticals tend to be preferentially located in the 
low-shear environment while the latest-type spirals tend to be preferentially 
located in the high-shear environment.

Nevertheless, it has to be noted that the galaxy-shear correlation signal 
is found to be statistically significant only in bins containing E and SIV 
galaxies. Therefore, when the results over all six bins are considered 
the null hypothesis of no galaxy-shear correlation is still quite acceptable. 
In the given density range, the null hypothesis is found to be rejected at 
only the $\sim 10\%$ confidence level when the results over all six bins are 
considered. Since we consider only those galaxies at similar densities, 
the sample size, $N_{g}$, is quite small, resulting in large statistical 
errors. It is also worth mentioning that neither in the highly underdense 
region with $\delta <-0.5$ nor in the highly overdense region with 
$\delta >1.0$, no signal of galaxy-shear correlation is found since 
the degree of density-shear correlations is too high to be removed in 
these regions.

\section{DISCUSSION AND CONCLUSION}

By measuring observationally the mean ellipticities of large-scale structures 
where the nearby galaxies are embedded as a function of galaxy morphological 
types, we have tentatively detected an independent relationship between galaxy 
morphology and shear of environment: In the mildly overdense environment with 
large scale dimensionless overdensity of $0.05\le\delta\le1.06$, the 
ellipticals are found to prefer the low-shear region; 
In the mildly underdense environment with large scale density of 
$-0.3\le\delta\le-0.1$, the latest-type spirals (Scd-Sm) are found to prefer 
the high-shear region. No correlation signal, however, is found either in the 
highly overdense ($\delta> 1$) or in the highly underdense environment 
($\delta<-0.5$).

This observational results may be explained by the dependence of halo 
formation epochs on the shear of large-scale environment. It has been known 
that the formation epochs of galactic halos depend not only on their mass but 
also on the density of large-scale environment \citep{gao-etal05}.  
Here we argue that there is some observational evidence to support the 
hypothesis that the formation epochs of halos also depend on the shear of the 
large-scale environment. In the high-shear environment the halos formed 
relatively early without growing at late time due to strong tidal disruption 
from the surrounding matter distribution. Thus, the late-type spirals are 
likely to be located in this high-shear environment. Meanwhile in the 
low-shear environment the halos formed relatively recently since there is 
no strong tidal distribution, having grown at late times through hierarchical 
merging. Hence, the ellipticals are likely to be found in this low-shear 
environment. In the highly overdense regions, however, even in case that 
there is tidal disruption from the surrounding matter distribution, the   
other stronger effects of environmental processes should mask the 
shear-dependence of galaxy properties. Likewise in the highly underdense 
regions, the tidal field is too weak to produce any significant 
galaxy-shear correlation. It will be interesting to test against N-body 
simulations whether or not the formation epochs of galactic halos with 
similar mass at similar density is a function of the ellipticity of 
large-scale dark matter distribution. Our future work is in this direction.

It should be noted here that our results suffer from low statistical 
significance due to the small sample size. We have seen only a marginal effect 
of the large-scale shear for the latest-type spirals (Scd-Sm). Meanwhile no 
signal has been found for the Sb-Sc galaxies even though both types of 
galaxies occupy similar regions. To remove the effect of the stronger 
galaxy-density correlations, we had to constrain the overdensities of 
galactic regions, which resulted in large errors.  The null hypothesis that 
there is no galaxy-shear correlation over the six bins is found to be rejected 
at only the $10\%$ level due to large errors. Given our result, it will 
require a sample of more than $100,000$ galaxies to test the hypothesis 
at $>90\%$ confidence level. This line of investigation will provide new 
insight into the formation and evolution of galaxies in a filamentary 
cosmic web.

\acknowledgments 

We are very grateful to P. Erdogdu for providing us the 2MRS density field.
We also thank an anonymous referee who helped us improve significantly the 
original manuscript by making many helpful suggestions. This work is supported 
by the Korea Science and Engineering Foundation (KOSEF) grant funded by the 
Korean Government (MOST, NO. R01-2007-000-10246-0). 

The 2MRS density field was constructed by \citet{erd-etal06} by making  
use of data products from the Two Micron All Sky Survey, which is a joint  
project of the University of Massachusetts and the Infrared Processing and 
Analysis Center/California Institute of Technology, funded by the National 
Aeronautics and Space Administration and the national Science Foundation, 
and the NASA/IPAC Extragalactic Database (NED) which is operated by the Jet 
Propulsion Laboratory, California Institute of Technology, under contract 
with the National Aeronautics and Space Administration and the SIMBAD 
database, operated at CDS, Strasbourg, France.

\clearpage
\begin{deluxetable}{cccc}
\tablewidth{0pt}
\setlength{\tabcolsep}{5mm}
\tablehead{Subsample & Class & Type & $N_{g}$ }
\tablecaption{}
\startdata   
E & Ellipticals & cE,E0,E0-1,E+ & $1008$ \\ 
L & Lenticulars & S0$^{-}$,S0$^{o}$,S0$^{+}$   & $2532$ \\ 
SI & Spirals & S0a,Sa  & $1761$ \\ 
SII & Spirals & Sab,Sb & $3175$  \\ 
SIII & Spirals & Sbc,Sc & $4507$ \\ 
SIV  & Spirals & Scd,Sd,Sdm,Sm & $2679$  \\ 
\enddata
\label{tab:type}
\end{deluxetable}
\clearpage
\clearpage
\clearpage
\clearpage
\clearpage
\clearpage
\clearpage
\end{document}